\documentclass[prl,twocolumn,preprintnumbers,superscriptaddress,showpacs,amsmath,amssymb]{revtex4}
\usepackage{graphicx}
\usepackage{hyperref}
\usepackage{amsmath}

\begin{document}


\title{Three-dimensional, self-similar, light in unstable canonical optical resonators}

\author{Steven W.\ Miller}
\affiliation{School of Physics \& Astronomy, College of Science \& Engineering, University of Glasgow,Glasgow G12~8QQ, United Kingdom}

\author{John Nelson}
\affiliation{APC, UMR 7164, Universit\'{e} Paris 7 Denis Diderot, 10 rue Alice Domon et L\'{e}onie Duquet, 75025 Paris Cedex 13, France}

\author{Johannes Courtial}
\email{Johannes.Courtial@glasgow.ac.uk}
\affiliation{School of Physics \& Astronomy, College of Science \& Engineering, University of Glasgow,Glasgow G12~8QQ, United Kingdom}

\date{\today}

\begin{abstract}
The eigenmodes of unstable canonical optical resonators possess fractal structure in their transverse intensity cross-sections [Karman \textit{et al.}, Nature \textbf{402}, 138Ê(1999)].
In one particular plane, the magnified self-conjugate plane, this structure can be explained in terms of a combination of imaging and diffraction [Courtial and Padgett, PRL \textbf{85}, 5320 (2000)].
Here we show that this combination of imaging and diffraction simultaneously occurs in the longitudinal direction, resulting in three-dimensional self-similar fractal structure around the centre of the magnified self-conjugate plane.
\end{abstract}




\maketitle

\newcommand{\rmi}{\mathrm{i}}
\newcommand{\rmd}{\mathrm{d}}
\newcommand{\bi}[1]{\mathbf{#1}}

\noindent
\textit{Introduction.}
Light can be fractal.
The (dark) vortex lines in random light fields have fractal scaling properties \cite{OHolleran-et-al-2008}.
Light's spatial (and spectral \cite{Lehman-Garavaglia-1999}) 
distribution can be directly made fractal by interaction with a fractal object, for example by emitting it from a fractal antenna \cite{Werner-Ganguly-2003},
by passing it through a fractal aperture \cite{Lehman-2001,Saavedra-et-al-2003},
or by resonating it in a resonator that contains a fractal scatterer \cite{Takeda-et-al-2004}.
Perhaps more surprisingly, the light field behind a (non-fractal) Ronchi grating illuminated by a uniform plane wave evolves, on propagation, into a fractal~\cite{Berry-Klein-1996a}.

Successive round trips through unstable canonical resonators also result in fractal transverse intensity distributions;
the lowest-loss eigenmodes of such resonators therefore have fractal intensity cross-sections \cite{Karman-Woerdman-1998,Karman-et-al-1999}.
A number of authors studied the properties (fractal dimension etc.) of these intensity distributions in planes where the intensity distribution is a statistical fractal \cite{Karman-Woerdman-1998,Karman-et-al-1999,Berry-2001,New-et-al-2001,Yates-New-2002,Loaiza-2005}.
Upon magnification, statistical fractals look like the same \emph{type} of pattern, but not actually the same pattern.
Natural examples of statistical fractals include many clouds, mountains, and Lichtenberg figures such as lightning.
Like in all physical fractals, the range of length scales over which this scaling behaviour holds (the scaling range) is limited \cite{Avnir-et-al-1998}, here by diffraction.

\begin{figure}
\centering
\includegraphics{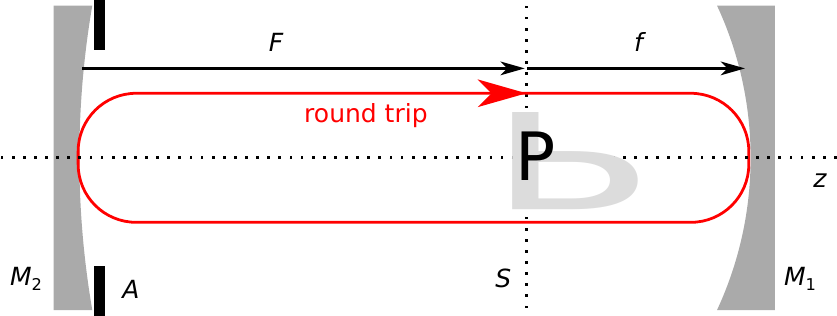}
\caption{\label{resonator-figure}
Imaging inside an unstable canonical resonator.
The two spherical mirrors, $M_1$ (focal length $f$) and $M_2$ (focal length $F$), perform geometric imaging.
$S$ is the magnified self-conjugate plane; the $z$ axis is chosen to coincide with the optical axis.
Three-dimensional imaging during one round trip is indicated by an object in the shape of the letter ``P'' (shown in black) and its image, which looks like a horizontally elongated letter ``b'' (shown in grey):
the ``P'' has turned into a ``b'' because the transverse magnification, $M$, is negative, and so the image of the ``P'' is upside-down;
the ``b'' is horizontally elongated because the longitudinal magnification, $M_l$, is positive and its magnitude is greater than that of the transverse magnification.
$A$ is an aperture.
The figure is drawn for a confocal resonator with $M = -2$ and $M_l = +4$.}
\end{figure}

\begin{figure}
\begin{center} \includegraphics{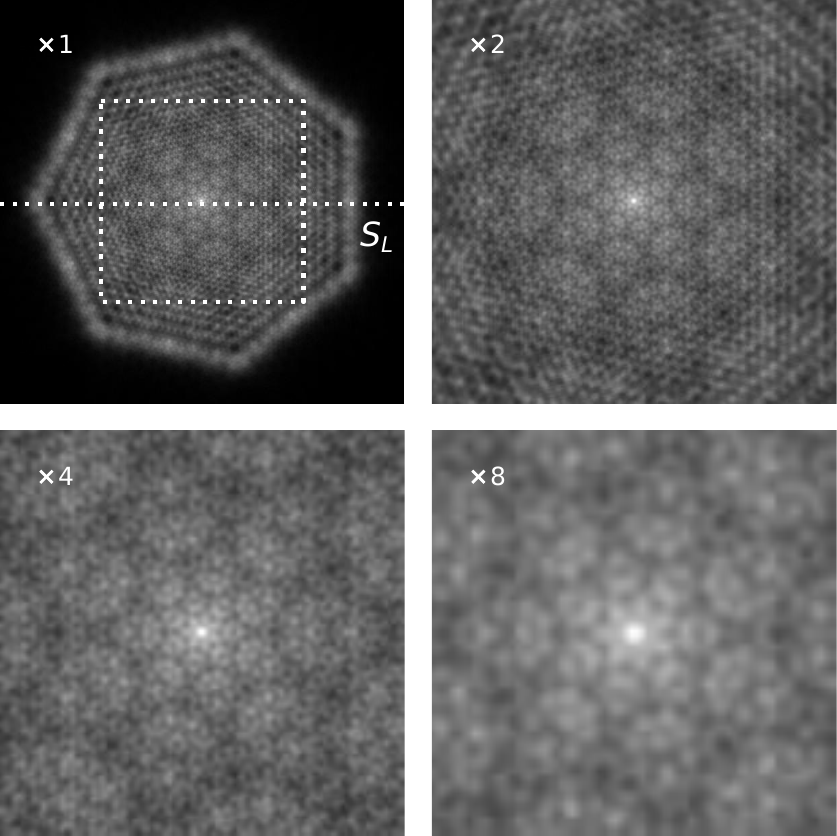} \end{center}
\caption{\label{transverse-intensity-distribution-3D-resonator-figure}Self-similarity of the lowest-loss eigenmode's intensity distribution in the magnified self-conjugate plane, $S$.
The frames show the intensity after 20 round trips, starting with a uniform plane wave.
The self-similarity of the pattern is demonstrated by showing its centre at different magnifications ($\times 2$, $\times 4$, $\times 8$), resulting in a similar pattern (rotated by $180^\circ$ after each magnification by a factor 2 due to the resonator's transverse magnification, $M$, being negative);
the dotted white square in the centre of the frame marked $\times 1$ shows the outline of the area shown in the next frame.
The horizontal dotted line is the orthographic projection of the lateral self-conjugate plane $S_L$ (see Fig.\ \ref{lateral-intensity-distribution-3D-resonator-figure}).
The figure is calculated for light of wavelength $\lambda = 632.8$\,nm in a resonator of the type shown in Fig.\ \ref{resonator-figure} with $F=16.5$\,cm, $f=8.25$\,cm, $M = -2$, and a seven-sided polygonal aperture of radius 2.4\,mm.
The beam's transverse cross-sections were represented by a $1024 \times 1024$ array of complex numbers sampled over a physical area of size 1\,cm$\times$1\,cm.
Like all simulations in this paper, this simulation was performed using the open-source wave-optics simulator Young TIM~\cite{Leavey-Courtial-2013}.
}
\end{figure}

Unstable canonical resonators contain a special plane, the magnified self-conjugate plane (see Fig.\ \ref{resonator-figure}), in which the intensity distribution is a self-similar fractal \cite{Courtial-Padgett-2000b}.
An example is shown in Fig.\ \ref{transverse-intensity-distribution-3D-resonator-figure}.
After magnification by a characteristic scaling factor, a part of a self-similar pattern looks the same as the pattern as a whole.
Suitable choice of the resonator parameters can even lead to intensity distributions closely related to classic fractals such as the Weierstrass-Mandelbrot function \cite{Courtial-Padgett-2000b}, the Sierpinski gasket, and the Koch snowflake~\cite{Watterson-et-al-2003}.

Here we investigate the three-dimensional (3D) intensity distribution in unstable canonical resonators.
We find that, around the centre of the magnified self-conjugate plane, this intensity distribution is a self-similar 3D fractal, albeit with different transverse and longitudinal characteristic scaling factors.

We will explain the emergence of this structure in terms of the properties of the resonator.
When not considered in the context of resonators, the existence of 3D self-similar fractal light fields is surprising:
the 3D intensity distribution of any light field is fully determined by any transverse cross-section, and so the lowest-loss eigenmode is fully determined by its cross-section in the magnified self-conjugate plane.
The existence of self-similar transverse cross-sections whose corresponding beams --- their 3D diffraction patterns --- are also self-similar is far from obvious.

\textit{Transverse fractals.}
We start by reviewing the mechanism that leads to self-similar fractal structure in an unstable canonical resonator's magnified self-conjugate plane.
Without loss of generality, we restrict ourselves to confocal resonators, as these are particularly simple but at the same time representative of all canonical unstable resonators (with the same round-trip magnification, $M$, and the same Fresnel number \cite{Siegman-1986}).
Fig.\ \ref{resonator-figure} shows such a resonator.


In a canonical resonator, each mirror is spherical and so images like a lens, but in reflection.
During one round trip, i.e.\ reflection off both mirrors, the image produced by the first mirror is imaged again by the second mirror; Ref.\ \cite{Nelson-et-al-2008} uses raytracing to visualise a few effects related to this imaging.
In stable resonators this imaging explains the eigenmodes' structural stability \cite{Forrester-et-al-2002}.
In unstable canonical resonators, one round trip images two ``self-conjugate'' planes back to their original positions, one with (transverse) magnification $M$ ($|M| \geq 1$), the other with magnification $1/M$ \cite{Courtial-Padgett-2000b}.
The former is the magnified self-conjugate plane, $S$, the latter the de-magnified self-conjugate plane, $s$.
In a confocal resonator, these planes are a focal distance on either side of the two mirrors (see Fig.\ \ref{resonator-figure}), and so the field these planes forms a Fourier pair.
Every round trip through the resonator stretches the intensity distributions in the planes $S$ and $s$ by a factor $M$ and $1/M$, respectively.

Any apertures in the resonator simply add some diffractive ``decoration'' to this image.
After a number of round trips, the pattern does not change any longer between successive round trips, which means the field has settled into an eigenmode.
In our case, the lowest-loss eigenmode is reached after approx.\ 20 round trips.
Once the eigenmode has formed, the decoration pattern is the same during successive round trips.
Once added, it gets magnified with the rest of the intensity distribution, which results in the presence of the decoration pattern in a number of sizes:
the pattern added during the most recent round trip is at the original size;
that added during the previous round trip is magnified by $M$;
that added two round trips ago is magnified by $M^2$;
and so on.
The presence of a pattern on such a cascade of length scales is a hallmark of self-similarity.
The mechanism outlined above is called the monitor-outside-a-monitor effect (MOM effect), named so because of analogies with video feedback~\cite{Courtial-et-al-2001,Leach-et-al-2003}.
Fig.\ \ref{transverse-intensity-distribution-3D-resonator-figure} demonstrates the self-similarity of an example of an eigenmode's intensity cross-section in the plane $S$.

\begin{figure}
\begin{center} \includegraphics{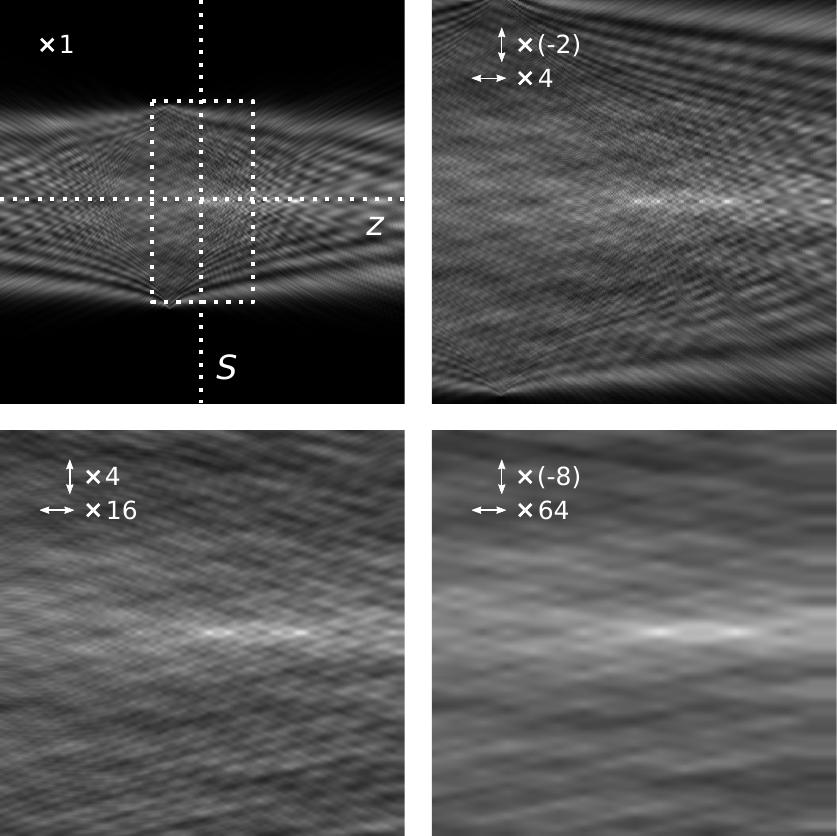} \end{center}
\caption{\label{lateral-intensity-distribution-3D-resonator-figure}Self-similarity of the intensity distribution in a lateral self-conjugate plane, $S_L$, which contains the optical axis and intersects the plane $S$ horizontally in Fig.\ \ref{transverse-intensity-distribution-3D-resonator-figure}.
The vertical dotted line is the orthographic projection of the plane $S$.
Vertically, the plots are centred on the optical axis, $z$.
The beam is the same as that shown in Fig.\ \ref{transverse-intensity-distribution-3D-resonator-figure}.
After each magnification horizontally by a factor 4 and vertically by $-2$, the pattern looks similar again, which is shown for different magnifications.
The dotted box in the centre of the frame marked $\times 1$ marks the outline of the area shown in the next frame ($\times (-2)$ vertically, $\times 4$ horizontally).
The $\times 1$ frame represents a physical area of size 2\,m (horizontally) by 10\,mm (vertically).}
\end{figure}

\begin{figure}
\begin{center} \includegraphics{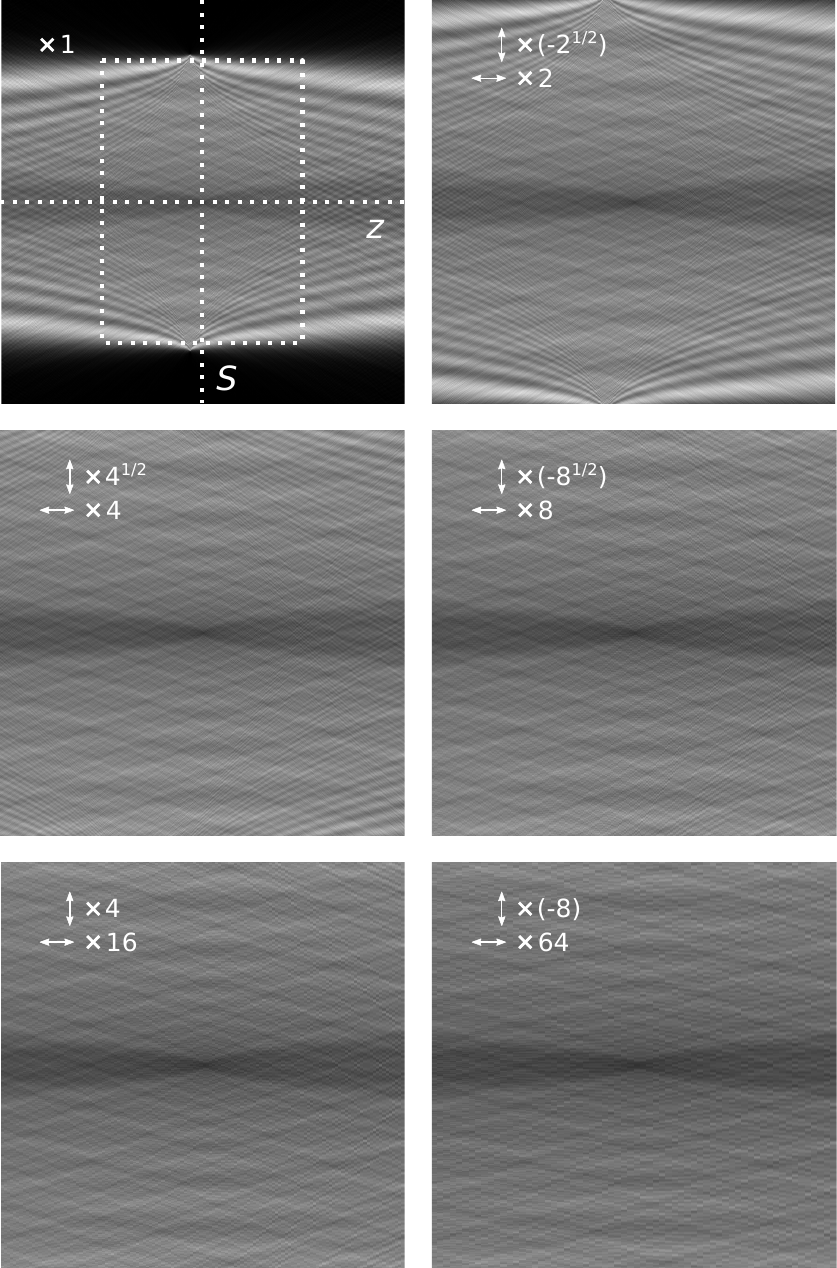} \end{center}
\caption{\label{lateral-intensity-distribution-2D-resonator-figure}Self-similarity of the intensity distribution in the lateral plane of a strip resonator of the type shown in Fig.\ \ref{resonator-figure}.
The different frames show the centre of the intensity distribution, successively magnified by a factor $M$ in the vertical direction and by $M^{2}$ in the horizontal direction.
The $\times 1$ frame represents a physical area of size 20\,m (horizontally) by 2.82\,cm (vertically), centred on the magnified self-conjugate plane and the optical axis in the horizontal and vertical direction, respectively.
The dotted box shown in the top left frame outlines the area shown in the top right frame.
The figure was calculated for light of wavelength $\lambda = 632.8$\,nm, resonator parameters $F = 70.7$\,cm and $f = 50$\,cm, and the aperture $A$ was a slit of width 2.08\,cm.
The beam cross-sections were represented on 4096-element array of complex numbers, representing a physical width 4\,cm.
}
\end{figure}

\textit{3D fractals.}
For the same resonator, Fig.\ \ref{lateral-intensity-distribution-3D-resonator-figure} shows a lateral intensity distribution around the centre of the self-conjugate plane $S$.
This lateral intensity distribution shows some signs of self-similarity:
if the pattern is stretched by $M$ in the direction representing the transverse direction, and by a factor $M^2$ in the longitudinal direction, the pattern's centre (which is the point where the plane $S$ intersects the resonator's optical axis) looks similar to what it was before magnification.
This self-similarity can be seen much clearer in Fig.\ \ref{lateral-intensity-distribution-2D-resonator-figure}, which was calculated for a strip resonator, i.e.\ a resonator that is invariant in one transverse direction.
It can therefore be treated as a 2D resonator with only one transverse direction, which means that, along that transverse direction, the transverse field can be represented by a much greater number of complex numbers without increasing memory or complexity requirements.
This in turn allows an increase in the Fresnel number by increasing the aperture size, resulting in a lateral intensity cross-section with significantly more detail.

\begin{figure}
\begin{center} \includegraphics{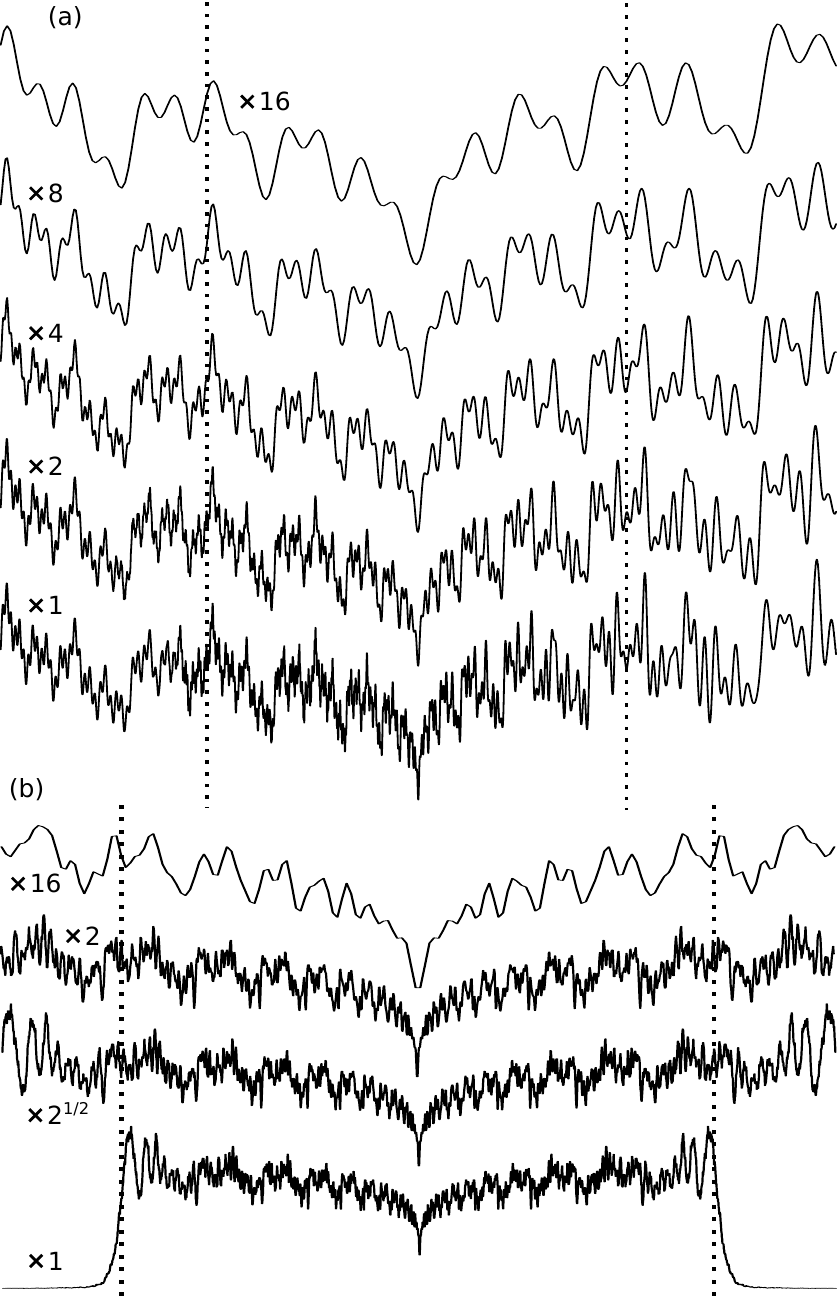} \end{center}
\caption{\label{line-intensity-distributions-2D-resonator-figure}(a)~Axial and (b)~transverse intensity cross-section through the field around the self-conjugate point at the centre of the plane $S$ in the strip resonator from Fig.\ \ref{lateral-intensity-distribution-2D-resonator-figure}.
Like in Figs \ref{transverse-intensity-distribution-3D-resonator-figure}, \ref{lateral-intensity-distribution-3D-resonator-figure} and \ref{lateral-intensity-distribution-2D-resonator-figure}, the self-similarity is demonstrated by successive magnifications, each of which stretches the part of the curve between the vertical dotted lines to the full width.
The width of the curves marked $\times 1$ represent a physical length of 2\,m (a) and 2.08\,cm (b).
The intensity range represented by the different curves has been adjusted so that corresponding features in the curves have roughly the same vertical size.}
\end{figure}

For that same strip resonator, Fig.\ \ref{line-intensity-distributions-2D-resonator-figure} compares the intensity cross-sections along the transverse direction in the plane $S$ with that along the resonator's optical axis.
The intensity cross-section along the optical axis is not symmetrical with respect to the position of the plane $S$, whereas that in the plane $S$ is symmetric with respect to the position of the optical axis.
But the self-similarity properties of these two curves are strikingly similar.

This observation can be explained as follows.
Spherical mirrors (and lenses) image not only any transverse plane into a corresponding transverse plane, they image any point into a corresponding point.
For light initially travelling to the right in the resonator shown in Fig.\ \ref{resonator-figure}, any lateral plane that includes the optical axis is being imaged into itself, as is the magnified self-conjugate plane $S$; no other planes are being imaged into themselves.
One point is imaged into itself (``self-conjugate point''), namely the intersection of the self-conjugate plane $S$ with the optical axis.
The volume around this point is also imaged into itself, but the image is distorted as the longitudinal and transverse magnifications are different (the longitudinal magnification is the square of the transverse magnification) and both change with position.
(Similar statements are true for light initially traveling to the left, but we do not consider these here.)
Close to the self-conjugate point, the longitudinal magnification is constant.
This imaging of the volume around the centre of the plane $S$ is indicated in Fig.\ \ref{resonator-figure}.

As before, the effect of any apertures in the system is the addition of a diffractive decoration pattern, which is now 3D.
In a 3D extension of the MOM effect, this pattern gets added to the field during each round trip and magnified during each subsequent round trip, again resulting in its presence on a cascade of length scales, complicated and enriched by the different characteristic stretch factors in the longitudinal and transverse directions.

\textit{Conclusions.}
We have found three-dimension self-similar fractal light in optical resonators.
We explain this structure in terms of a 3D monitor-outside-a-monitor effect:  the interplay between the 3D imaging properties of unstable resonators and diffraction.

\end{document}